\documentclass[11pt,a4paper]{article}

\usepackage[a4paper,margin=25mm]{geometry}
\usepackage{lmodern}
\usepackage[T1]{fontenc}
\usepackage{amsmath,amssymb,amsthm,mathtools}
\usepackage{booktabs,tabularx}
\usepackage{microtype}
\usepackage{enumitem}
\usepackage[hidelinks]{hyperref}

\setlist{itemsep=1pt,topsep=3pt}
\hypersetup{
  pdftitle={Who Observes Whom: Network Rationalizability and Its Complexity},
  pdfauthor={Mojtaba FayazBakhsh and MohammadAmin Fazli}
}

\newtheorem{theorem}{Theorem}[section]

\newtheorem{proposition}[theorem]{Proposition}

\theoremstyle{definition}
\newtheorem{definition}[theorem]{Definition}
\newtheorem{example}[theorem]{Example}
\theoremstyle{remark}
\newtheorem{remark}[theorem]{Remark}

\newcommand{\BR}{\mathrm{BR}}
\newcommand{\BRp}{\mathrm{BR}^{\mathrm{pure}}}
\newcommand{\supp}{\mathrm{supp}}
\newcommand{\Nbr}{N}
\newcommand{\cNbr}{\overline{N}}
\newcommand{\conv}{\mathrm{conv}}
\newcommand{\Rat}{\mathcal{R}}
\newcommand{\RatB}{\mathcal{R}^{\beta}}
\newcommand{\RatP}{\mathcal{R}^{\mathrm{pure}}}
\newcommand{\RatCor}{\mathcal{R}^{\mathrm{cor}}}
\newcommand{\Op}{\mathcal{T}}
\newcommand{\OpB}{\mathcal{T}^{\beta}}
\newcommand{\ER}{\exists\mathbb{R}}
\newcommand{\PPAD}{\mathsf{PPAD}}
\newcommand{\FIXP}{\mathsf{FIXP}}
\newcommand{\NP}{\mathsf{NP}}
\newcommand{\Poly}{\mathsf{P}}
\newcommand{\NE}{\mathrm{NE}}

\title{\textbf{Who Observes Whom:\\ Network Rationalizability and Its Complexity}}
\author{Mojtaba FayazBakhsh \and MohammadAmin Fazli}
\date{November 2026}

\begin{document}

\maketitle

\begin{abstract}
\noindent
Nash equilibrium and rationalizability sit at opposite informational extremes: the former presumes that each player knows all opponents' strategies, the latter that players reason from common knowledge of rationality alone. When visibility is partial and asymmetric, we study the solution concepts induced by a directed \emph{observation network} $G$ in which an edge $i\to j$ means that $i$ observes $j$'s strategy. A set of profiles is \emph{self-sustaining} if every profile in it can be justified, for every player, by conjectures from the same set that agree with everything the player observes. We consider \emph{point} conjectures (single profiles) and \emph{belief} conjectures (distributions over profiles). Both range from Nash equilibrium on the complete network to point- and correlated rationalizability on the empty network, and shrink monotonically as links are added.

Network topology governs both behavior and complexity. A predicted profile can be found in polynomial time on the empty network, whereas a single link makes search $\mathsf{PPAD}$-hard: under one-sided two-player monitoring, the observed player's strategies are exactly her Nash equilibrium strategies. On hierarchical networks (transitive tournaments), membership reduces recursively to smaller games and admits polynomial-size certificates, placing verification in $\exists\mathbb{R}$, and in $\mathsf{P}$ for two players. On the directed three-cycle, however, iterated elimination may never terminate under either variant, and under point conjectures interior predictions may admit only infinite witness sets, a pathology that belief conjectures avoid in the same game. Finally, when only realized pure actions are observed, the solution set is guaranteed to be non-empty exactly when the network has no edges, and is computable in time $O(n|A|^2)$.
\end{abstract}

\medskip
\noindent\textbf{Keywords:} rationalizability, conjectural equilibrium, observation networks, partial monitoring, $\PPAD$, existential theory of the reals.

\noindent\textbf{JEL Classification:} C72, D82, D83, D85.

\section{Introduction}

Two benchmark solution concepts frame non-cooperative game theory. Under the epistemic reading of \emph{Nash equilibrium} \cite{nash1950equilibrium,aumann1995epistemic}, every player correctly knows the strategies chosen by the others and best-responds to them. \emph{Rationalizability} \cite{bernheim1984rationalizable,pearce1984rationalizable} drops all such knowledge and characterizes the implications of common knowledge of rationality and payoffs \cite{tan1988bayesian}. Many environments lie between these extremes. In supply chains, organizational hierarchies, markets with local intelligence and multi-agent systems, each agent observes the behavior of particular peers and remains uninformed about everyone else. This paper asks what rational players do in such environments and how the structure of who observes whom shapes both the predictions and the computational cost of deriving them.

We model the observation structure as a directed network $G$ on the players and let player $i$ observe the strategies of her out-neighbors $\Nbr_i(G)$. Under common knowledge of rationality, payoffs and $G$, a set of profiles $\mathcal{S}$ is \emph{self-sustaining} if every $\sigma\in\mathcal{S}$ can be justified, for each player $i$, by conjectures drawn from $\mathcal{S}$ that coincide with $\sigma$ on $i$'s \emph{closed} neighborhood $\cNbr_i(G)=\Nbr_i(G)\cup\{i\}$. The largest self-sustaining set is the prediction. We consider two forms of justification. Under \emph{point conjectures}, $i$ justifies $\sigma_i$ by a single profile $\sigma'\in\mathcal{S}$; this yields $\Rat(\Gamma,G)$. Under \emph{belief conjectures}, $i$ may hold a probability distribution over such profiles; this yields $\RatB(\Gamma,G)$. The belief variant is the natural specialization of rationalizable conjectural equilibrium \cite{rubinstein1994rationalizable,esponda2013rationalizable} to signals that reveal neighbors' strategies; the point variant is its refinement in which each player entertains a single hypothesis about the state of play.

Both variants equal Nash equilibrium on the complete network $K_n$. On the empty network, $\Rat$ is Bernheim's point-rationalizability \cite{bernheim1984rationalizable} and $\RatB$ is correlated rationalizability \cite{brandenburger1987rationalizability}, which for two players is Bernheim--Pearce rationalizability. For every game and every network,
\begin{equation}\label{eq:chain}
  \NE(\Gamma)\;\subseteq\;\Rat(\Gamma,G)\;\subseteq\;\RatB(\Gamma,G)\;\subseteq\;\RatCor(\Gamma),
\end{equation}
and both variants expand as links are removed. The general fixed-point properties of these sets are of the kind familiar from the conjectural-equilibrium literature, and we collect them in Section~\ref{sec:model}. Our contribution is to show how the \emph{topology} of the network determines the predictions and their complexity:

\begin{enumerate}[label=(\arabic*)]
  \item \textbf{One link suffices for hardness.} On the empty network a predicted profile can be found in polynomial time. With a single one-sided link between two players, the observed player's strategies under either variant are exactly her Nash equilibrium strategies, while the observer best-responds to them (Theorem~\ref{thm:single-edge}). Consequently, finding a predicted profile is $\PPAD$-hard on every network with at least one edge, even when all players have non-trivial action sets (Theorem~\ref{thm:hardness}). Monitoring thus constrains the monitored: being observed removes a player's freedom to be mis-predicted.
  \item \textbf{Hierarchies admit compact certificates.} When the network is a transitive tournament (equivalently, both $G$ and its complement are acyclic), membership reduces recursively to smaller games. Under point conjectures this yields witness trees of $2^n-n-1$ strategy vectors, linear in the size of the normal form, and under belief conjectures certificates of polynomial size. Verification is therefore in $\ER$, and in $\Poly$ for two players (Theorem~\ref{thm:tournament}).
  \item \textbf{Cycles break finiteness.} On the directed three-cycle we construct $2\times2\times2$ games in which every round of elimination removes interior profiles, so the process never terminates, under both variants (Theorem~\ref{thm:doubling}). In a dual game, every interior profile is predicted but under point conjectures only infinite self-sustaining sets witness it, whereas belief conjectures admit witnesses with $27$ profiles (Theorem~\ref{thm:halving}). Non-termination thus persists under beliefs, while in this game infinite witnesses arise only from point conjectures.
  \item \textbf{Ex-post monitoring.} When realized pure actions rather than strategies are observed, the solution set is non-empty for every game if and only if the network has no edges, and it can be computed exactly in time $O(n|A|^2)$ for every network (Section~\ref{sec:expost}).
\end{enumerate}
Table~\ref{tab:summary} summarizes the results.

\begin{table}[t]
\centering
\caption{Summary of results for an $n$-player game with pure action profiles $A=\prod_iA_i$. Unless stated otherwise, results hold for both point ($\Rat$) and belief ($\RatB$) conjectures.}
\label{tab:summary}
\vspace{2pt}
\small
\begin{tabularx}{\textwidth}{>{\raggedright\arraybackslash}p{3.1cm} >{\raggedright\arraybackslash}p{2.5cm} >{\raggedright\arraybackslash}X}
\toprule
\textbf{Network} & \textbf{Search} & \textbf{Structure and verification} \\
\midrule
Empty & $\Poly$ & $\Rat$: point-rationalizability, finite union of faces, finite stabilization (Thm.~\ref{thm:empty}). $\RatB$: correlated rationalizability (Thm.~\ref{thm:endpoints}). \\
\addlinespace[2pt]
One-sided edge, $n=2$ & $\PPAD$-complete & $\{(\sigma_1,\sigma_2):\sigma_2\in\pi_2(\NE),\ \sigma_1\in\BR_1(\sigma_2)\}$; verification by one LP (Thm.~\ref{thm:single-edge}). \\
\addlinespace[2pt]
Any network, $E\neq\emptyset$ & $\PPAD$-hard & Hardness persists when $|A_i|\ge2$ for all $i$ (Thm.~\ref{thm:hardness}). \\
\addlinespace[2pt]
Transitive tournament & $\PPAD$-hard & Recursive characterization; certificates of size $2^n-n-1$ ($\Rat$) or polynomial ($\RatB$); verification in $\ER$, $\Poly$ for $n=2$ (Thm.~\ref{thm:tournament}). \\
\addlinespace[2pt]
Directed $3$-cycle & $\PPAD$-hard & Non-terminating elimination (Thm.~\ref{thm:doubling}); only infinite witness sets for $\Rat$ but finite ones for $\RatB$ in the same game (Thm.~\ref{thm:halving}). \\
\addlinespace[2pt]
Complete ($K_n$) & As Nash & Nash equilibrium; verification in $\Poly$. \\
\addlinespace[2pt]
Ex-post monitoring & $\Poly$ & Non-empty for all games iff $E=\emptyset$; full set in $O(n|A|^2)$ (Thms.~\ref{thm:expost-exist},~\ref{thm:expost-poly}). \\
\bottomrule
\end{tabularx}
\end{table}

\paragraph{Related literature.}
\emph{Rationalizability.} Bernheim \cite{bernheim1984rationalizable} introduced rationalizability together with point-rationalizability, in which conjectures are single strategy profiles, and showed that point-rationalizable strategies are rationalizable; Pearce \cite{pearce1984rationalizable} characterized rationalizability by iterated elimination of never-best responses, and Brandenburger and Dekel \cite{brandenburger1987rationalizability} studied the correlated version, which coincides with iterated strict dominance. Battigalli and Siniscalchi's $\Delta$-rationalizability \cite{battigalli2003rationalization} restricts first-order beliefs by exogenous information. In our setting the restriction is endogenous: what $i$ knows about $j$ is the strategy $j$ actually uses, which must itself be justified.

\emph{Conjectural equilibrium.} Conjectural equilibrium \cite{battigalli1987comportamento,battigalli1997conjectural}, subjective equilibrium \cite{kalai1995subjective}, self-confirming equilibrium \cite{fudenberg1993self}, Berk--Nash equilibrium \cite{esponda2016berk} and non-Nash steady states \cite{gilli1999non} require beliefs to be consistent with feedback about play. Rubinstein and Wolinsky \cite{rubinstein1994rationalizable} combined such consistency with rationalizability, and Esponda \cite{esponda2013rationalizable} developed the resulting framework for robust predictions. Our belief variant is the instance of this approach in which signals are projections of the strategy profile onto neighborhoods of a network. Rather than general signal structures, we study how graph-theoretic properties of the network (one-sided links, acyclicity, cycles) determine the predictions, their finiteness, and their complexity. In network formation, McBride \cite{mcbride2006imperfect} applies conjectural equilibrium under local monitoring, and Foerster, Mauleon and Vannetelbosch \cite{foerster2021shadow} study rationalizable conjectural pairwise stability when some links are hidden, showing that larger observation radii shrink the set of stable networks, a counterpart of the monotonicity in Theorem~\ref{thm:monotone}.

\emph{Epistemics and networks.} Aumann and Brandenburger \cite{aumann1995epistemic} show that mutual knowledge of conjectures yields Nash equilibrium; network rationalizability localizes such knowledge along the edges of $G$. In graphical games \cite{kearns2001graphical} and network games \cite{galeotti2010network} the network encodes payoff interactions; here payoffs are arbitrary and the network encodes information.

\emph{Complexity.} Computing a Nash equilibrium is $\PPAD$-complete for two players \cite{daskalakis2009complexity,chen2009settling}; exact $n$-player equilibria are $\FIXP$-complete \cite{etessami2010complexity}, and many decision problems about equilibria are $\NP$-hard \cite{gilboa1989nash} or $\ER$-complete \cite{schaefer2017fixed}. Two-player rationalizability, by contrast, reduces to iterated strict dominance and is solvable by linear programming. Our results locate the transition between these regimes in the network.

\section{Model}\label{sec:model}

\subsection{Games, networks and self-sustaining sets}

A finite game is $\Gamma=\langle N,(A_i)_{i\in N},(u_i)_{i\in N}\rangle$ with players $N=\{1,\dots,n\}$, finite action sets $A_i$, pure profiles $A=\prod_iA_i$ and payoffs $u_i:A\to\mathbb{R}$, extended multilinearly to mixed profiles $\sigma\in\Sigma=\prod_i\Delta(A_i)$. We write $\Sigma_{-i}=\prod_{j\neq i}\Delta(A_j)$, $\sigma_M$ for the restriction of $\sigma$ to $M\subseteq N$, and
\[
  \BR_i(\sigma_{-i})=\operatorname*{argmax}_{\tau_i\in\Delta(A_i)}u_i(\tau_i,\sigma_{-i}),\qquad
  \BRp_i(\sigma_{-i})=\operatorname*{argmax}_{a_i\in A_i}u_i(a_i,\sigma_{-i}).
\]
For a finitely supported distribution $\mu$ over $\Sigma$ we write $u_i(\tau_i,\mu)=\sum_{\sigma'}\mu(\sigma')u_i(\tau_i,\sigma'_{-i})$ and define $\BR_i(\mu)$ and $\BRp_i(\mu)$ accordingly. Since $u_i(\tau_i,\mu)$ is linear in $\tau_i$, a mixed strategy is a best response exactly when every action in its support is:
\begin{equation}\label{eq:support}
  \tau_i\in\BR_i(\mu)\iff\supp(\tau_i)\subseteq\BRp_i(\mu).
\end{equation}

An \emph{observation network} is a directed graph $G=(N,E)$ without self-loops; $(i,j)\in E$ means that $i$ observes the strategy of $j$. Let $\Nbr_i(G)=\{j:(i,j)\in E\}$ and $\cNbr_i(G)=\Nbr_i(G)\cup\{i\}$. For $\mathcal{S}\subseteq\Sigma$ and $\sigma\in\Sigma$, the profiles in $\mathcal{S}$ that player $i$ cannot distinguish from $\sigma$ form the \emph{fiber}
\[
  W_i(\sigma,\mathcal{S})=\{\sigma'\in\mathcal{S}:\ \sigma'_{\cNbr_i(G)}=\sigma_{\cNbr_i(G)}\}.
\]

\begin{definition}[Network rationalizability]\label{def:rat}
Fix $\Gamma$ and $G$. A set $\mathcal{S}\subseteq\Sigma$ is \emph{self-sustaining under point conjectures} if for every $\sigma\in\mathcal{S}$ and $i\in N$ there is $\sigma'\in W_i(\sigma,\mathcal{S})$ with $\sigma_i\in\BR_i(\sigma'_{-i})$. It is \emph{self-sustaining under belief conjectures} if for every $\sigma\in\mathcal{S}$ and $i\in N$ there is a finitely supported distribution $\mu$ on $W_i(\sigma,\mathcal{S})$ with $\sigma_i\in\BR_i(\mu)$. The \emph{network-rationalizable sets} $\Rat(\Gamma,G)$ and $\RatB(\Gamma,G)$ are the unions of all self-sustaining sets of the respective kind.
\end{definition}

Point conjectures are beliefs concentrated on a single profile, so every set that is self-sustaining under point conjectures is self-sustaining under belief conjectures, and $\Rat(\Gamma,G)\subseteq\RatB(\Gamma,G)$. By Carath\'eodory's theorem applied to the vector of expected payoffs in $\mathbb{R}^{A_i}$, a justifying belief can always be taken to have at most $|A_i|+1$ atoms.

\paragraph{Interpretation.}
A conjecture is a state of the world that player $i$ cannot rule out: it agrees with everything she observes and with her own choice, and, because it lies in $\mathcal{S}$, it is itself coherent from every player's point of view. Iterating this requirement expresses common knowledge of rationality, payoffs and the network. The two variants differ in how players treat uncertainty about what they do not see. Under belief conjectures, $i$ may hedge across several coherent states; under point conjectures she commits to one hypothesis. A conjectured mixed strategy is then a strategy its owner might actually be using, so it can put weight on several actions only if they are \emph{simultaneously} optimal for that owner. This restriction is what separates point- from ordinary rationalizability on the empty network (Example~\ref{ex:separation}).

An edge $(i,j)$ means that $i$ observes the mixed strategy $\sigma_j$ (Section~\ref{sec:expost} treats observation of realized actions). Natural readings are Nash's mass-action interpretation \cite{nash1951noncooperative}, in which $\sigma_j$ is the distribution of actions in population $j$; delegated play through auditable randomized programs; and the view of mixed strategies as transparent intentions, which is exactly the informational assumption behind mixed Nash equilibrium, here imposed only along the edges of $G$. As in Nash equilibrium, a player evaluates deviations holding the strategies of others fixed: observation conveys information about what others do, not the power to influence it (Remark~\ref{rem:stackelberg}).

Including $i$ in $\cNbr_i(G)$ encodes that $i$ knows her own strategy and knows that those who monitor her see that strategy rather than a counterfactual one. It implies a \emph{self-witnessing} property used repeatedly below: every profile in the support of a justifying conjecture agrees with $\sigma$ on $i$'s coordinate. Dropping $i$ from her own neighborhood produces implausible predictions.

\begin{example}[Self-knowledge matters]\label{ex:irreflexive}
Let $A_1=\{T,B\}$, $A_2=\{L,R\}$, $E=\{(1,2)\}$, and $u(T,L)=(1,1)$, $u(T,R)=(0,1)$, $u(B,L)=(0,1)$, $u(B,R)=(1,0)$. If witnesses had to agree only on $\Nbr_i(G)$, the set $\{(T,L),(B,R)\}$ would be self-sustaining: in $(B,R)$, player~2 could justify $R$ by the witness $(T,L)$, in which she herself plays $L$ and player~1, observing $L$, plays $T$. Under Definition~\ref{def:rat}, player~2 knows that she plays $R$ and that player~1 sees it; hence player~1 plays $B$, against which $R$ is strictly worse.
\end{example}

\begin{example}[Partial consensus]\label{ex:three-player}
Let players 1 and 2 observe each other and player~3 be isolated. Player~3 has a constant payoff and selects whether players 1 and 2 play Chicken or Matching Pennies. Then $(\sigma_1,\sigma_2,\sigma_3)\in\Rat(\Gamma,G)$ if and only if $\sigma_1\in\BR_1(\sigma_2,\sigma'_3)$ and $\sigma_2\in\BR_2(\sigma_1,\sigma''_3)$ for some, possibly different, $\sigma'_3,\sigma''_3$. Players 1 and 2 are mutually consistent in the Nash sense but may disagree about the unobserved state of the world.
\end{example}

\begin{remark}[Coordinatewise convexification]\label{rem:product-hull}
One might instead allow $i$ to conjecture any $\sigma'$ with $\sigma'_{\cNbr_i}=\sigma_{\cNbr_i}$ and $\sigma'_j\in\conv(\pi_j(\mathcal{S}))$ for the other coordinates. This ignores what unobserved players themselves observe. With $E=\{(1,2)\}$ in Matching Pennies, it would let player~2 believe that player~1 mixes uniformly, although player~2 knows that player~1 sees her strategy and best-responds to it; every strategy of player~2 would survive. Belief conjectures over fibers avoid this, because each profile in the support of the belief remains coherent.
\end{remark}

\subsection{Fixed points and elimination}

Let $\Op_G(\mathcal{S})$ and $\OpB_G(\mathcal{S})$ be the sets of $\sigma\in\Sigma$ that admit, for every $i$, a point or belief conjecture in $W_i(\sigma,\mathcal{S})$ as in Definition~\ref{def:rat}. A set is self-sustaining exactly when $\mathcal{S}\subseteq\Op_G(\mathcal{S})$, respectively $\mathcal{S}\subseteq\OpB_G(\mathcal{S})$. Define elimination stages $\Rat_0=\Sigma$, $\Rat_{t+1}=\Op_G(\Rat_t)$, and analogously $\RatB_t$.

\begin{theorem}[Fixed-point structure]\label{thm:fixed-point}
For every game $\Gamma$, network $G$ and both variants:
\begin{enumerate}[label=(\roman*)]
  \item $\Rat(\Gamma,G)$ and $\RatB(\Gamma,G)$ are the greatest fixed points of $\Op_G$ and $\OpB_G$;
  \item the stages $\Rat_t$ and $\RatB_t$ are non-increasing, non-empty, compact and semi-algebraic;
  \item $\Rat(\Gamma,G)=\bigcap_t\Rat_t$ and $\RatB(\Gamma,G)=\bigcap_t\RatB_t$.
\end{enumerate}
\end{theorem}

\begin{proof}
Both operators are monotone, and the union of post-fixed points is a post-fixed point; Tarski's theorem gives (i). For (ii), with $k_i=|A_i|+1$, $\OpB_G(\mathcal{S})$ is the intersection over $i$ of the projections onto $\sigma$ of the sets of tuples $(\sigma,\sigma^1,\dots,\sigma^{k_i},\lambda)$ with $\sigma^m\in\mathcal{S}$, $\sigma^m_{\cNbr_i}=\sigma_{\cNbr_i}$, $\lambda\in\Delta_{k_i}$ and $\sum_m\lambda_mu_i(\sigma_i,\sigma^m_{-i})\ge\sum_m\lambda_mu_i(a_i,\sigma^m_{-i})$ for all $a_i$. If $\mathcal{S}$ is compact and semi-algebraic, so is this set; projection preserves compactness and, by the Tarski--Seidenberg theorem \cite{tarski1951decision}, semi-algebraicity. The point variant is the case $k_i=1$. Since $\Op_G(\mathcal{S})\subseteq\OpB_G(\mathcal{S})$, induction gives $\Rat_t\subseteq\RatB_t$. Monotonicity yields that the stages are non-increasing, and non-emptiness follows from $\NE(\Gamma)\subseteq\Rat_t\subseteq\RatB_t$ (Theorem~\ref{thm:monotone}). For (iii), the inclusion $\subseteq$ holds by induction. Conversely, if $\sigma\in\bigcap_t\RatB_t$, then for each $i$ the sets of justifying tuples with $\sigma^m\in\RatB_t$ are non-empty, compact and nested in $t$, so by Cantor's intersection theorem some tuple works for all $t$; its atoms lie in $\bigcap_t\RatB_t$. Hence $\bigcap_t\RatB_t$ is self-sustaining. The same argument applies to $\Rat$.
\end{proof}

\begin{theorem}[Monotonicity and existence]\label{thm:monotone}
If $E'\subseteq E$, then $\Rat(\Gamma,G)\subseteq\Rat(\Gamma,G')$ and $\RatB(\Gamma,G)\subseteq\RatB(\Gamma,G')$. Moreover $\Rat(\Gamma,K_n)=\RatB(\Gamma,K_n)=\NE(\Gamma)$, so $\emptyset\neq\NE(\Gamma)\subseteq\Rat(\Gamma,G)\subseteq\RatB(\Gamma,G)$ for every $G$.
\end{theorem}

\begin{proof}
Removing edges enlarges fibers, so any justification under $G$ remains valid under $G'$. On $K_n$ the fiber of $\sigma$ in any $\mathcal{S}\ni\sigma$ is $\{\sigma\}$, so the condition is $\sigma_i\in\BR_i(\sigma_{-i})$ for all $i$. Non-emptiness is Nash's theorem.
\end{proof}

We also record that pure profiles which are self-sustaining as pure profiles (Section~\ref{sec:expost}) remain so as degenerate mixed profiles: since $a_i\in\BRp_i(a'_{-i})$ if and only if $\delta_{a_i}\in\BR_i(\delta_{a'_{-i}})$,
\begin{equation}\label{eq:pure-in-mixed}
  \RatP(\Gamma,G)\subseteq\Rat(\Gamma,G)\subseteq\RatB(\Gamma,G).
\end{equation}

\begin{remark}[Support cells]\label{rem:cells}
By~\eqref{eq:support}, the condition for player $i$ at $\sigma$ depends on $\sigma_i$ only through $\supp(\sigma_i)$. On each of the finitely many cells on which the support profile $(\supp(\sigma_i))_i$ is constant, the operators are therefore described by a fixed system of polynomial inequalities linking the observed coordinates of $\sigma$ to the conjectured ones. Section~\ref{sec:cycle} shows that, on cyclic networks, iterating these finitely many rules can nonetheless require infinitely many steps.
\end{remark}

\section{The Empty Network}\label{sec:empty}

When $G=\emptyset$, fibers constrain only a player's own coordinate. We begin with a decomposition that holds on every network.

\begin{proposition}[Decomposition]\label{prop:cartesian}
For both variants:
\begin{enumerate}[label=(\roman*)]
  \item if $N_1,\dots,N_m$ are the weakly connected components of $G$, every stage and the limit are Cartesian products $\prod_r\pi_{N_r}(\cdot)$;
  \item if $\Nbr_i(G)=N\setminus\{i\}$, player $i$'s condition at any $\sigma$ in the set under consideration is $\sigma_i\in\BR_i(\sigma_{-i})$;
  \item if $k\in\Nbr_i(G)$ and $\Nbr_k(G)\subseteq\cNbr_i(G)$ (for instance, if $G$ is transitive), then every justification of $k$ at $\sigma$ is also a justification of $k$ at every profile in the support of a justification of $i$ at $\sigma$.
\end{enumerate}
\end{proposition}

\begin{proof}
(i) For $i\in N_r$, $\cNbr_i\subseteq N_r$; when $\mathcal{S}$ is a product, $i$'s condition constrains only $\sigma_{N_r}$, so both operators map products to products. Induction from $\Sigma$ and Theorem~\ref{thm:fixed-point}(iii) give the claim. (ii) The fiber is $\{\sigma\}$. (iii) Profiles in $i$'s fiber agree with $\sigma$ on $\cNbr_k\subseteq\cNbr_i$, hence have the same fiber for $k$.
\end{proof}

\begin{theorem}[Point conjectures on the empty network]\label{thm:empty}
\begin{enumerate}[label=(\roman*)]
  \item $\Rat(\Gamma,\emptyset)=\prod_iP_i$, where $(P_i)_i$ is the largest family with $P_i\subseteq\Delta(A_i)$ such that each $\sigma_i\in P_i$ is a best response to some $\sigma'_{-i}\in\prod_{j\neq i}P_j$. That is, $\Rat(\Gamma,\emptyset)$ is the set of point-rationalizable profiles \cite{bernheim1984rationalizable}.
  \item Each stage has the form $\prod_i\bigcup_{B_i\in\mathcal{B}_i^{(t)}}\Delta(B_i)$ for downward-closed families $\mathcal{B}_i^{(t)}$ of non-empty supports, and the stages stabilize after at most $\sum_i(2^{|A_i|}-1)$ rounds.
  \item Membership is decidable; for $n=2$ the limit families can be computed by linear programming in time polynomial in the size of the game times $2^{O(|A_1|+|A_2|)}$.
\end{enumerate}
\end{theorem}

\begin{proof}
(i) By Proposition~\ref{prop:cartesian}(i) the limit is a product, and for a product $\prod_jP_j$ a witness for $i$ is any $\sigma'$ with $\sigma'_i=\sigma_i$ and $\sigma'_{-i}\in\prod_{j\neq i}P_j$. (ii) By~\eqref{eq:support}, $\sigma_i$ survives a round if and only if the region $\{\sigma'_{-i}:\supp(\sigma_i)\subseteq\BRp_i(\sigma'_{-i})\}$ meets the product of the opponents' current sets, a condition that depends only on $\supp(\sigma_i)$ and that is inherited by subsets of supports. The families are non-increasing, and each round that changes something removes a support. (iii) A support $B_i$ survives if and only if, for some choice of surviving supports $B_j$ of the opponents, the system $\sigma'_j\in\Delta(B_j)$, $B_i\subseteq\BRp_i(\sigma'_{-i})$ is feasible. For $n=2$ this is a linear program; for $n\ge3$ it is a sentence of the existential theory of the reals.
\end{proof}

Under point conjectures, a mixed conjecture about $j$ may put weight only on actions that are simultaneously optimal for $j$. The next example shows that this can eliminate pure actions that are rationalizable, so the inclusion of point-rationalizability in rationalizability \cite{bernheim1984rationalizable} can be strict even on pure actions.

\begin{example}[Point- versus ordinary rationalizability]\label{ex:separation}
In the game of Table~\ref{tab:separation}, no pure action is strictly dominated: $T$, $B$ and $D$ are the unique best responses to $L$, $R$ and $\tfrac12L+\tfrac12R$, and $L$, $R$, $M$ are the unique best responses to $T$, $B$, $D$. Hence all pure profiles are rationalizable, and $\RatB(\Gamma,\emptyset)$ contains all of them by Theorem~\ref{thm:endpoints}. Under point conjectures, write $\sigma'_1=(\alpha_T,\alpha_D,\alpha_B)$ and $\tilde\alpha=\alpha_T+\tfrac12\alpha_D$; then player~2's payoffs from $L$, $M$, $R$ are $3\tilde\alpha$, $2$ and $3(1-\tilde\alpha)$. Both $L$ and $R$ are optimal only at $\tilde\alpha=\tfrac12$, where $M$ is strictly better. So every surviving strategy of player~2 has $q_L=0$ or $q_R=0$. Against $q_R=0$, $u_1(T)-u_1(D)=10-9q_L\ge1$, and symmetrically against $q_L=0$; hence $D$ is eliminated in the second round, and $\Rat(\Gamma,\emptyset)\subsetneq\RatB(\Gamma,\emptyset)$.
\end{example}

\begin{table}[htbp]
\centering
\caption{$D$ is rationalizable but not point-rationalizable.}
\label{tab:separation}
\vspace{2pt}
\begin{tabular}{l c c c}
\toprule
 & $L$ & $M$ & $R$ \\
\midrule
$T$ & $3,\ 3$ & $0,\ 2$ & $0,\ 0$ \\
$D$ & $2,\ 1.5$ & $-10,\ 2$ & $2,\ 1.5$ \\
$B$ & $0,\ 0$ & $0,\ 2$ & $3,\ 3$ \\
\bottomrule
\end{tabular}
\end{table}

Let $A_i^*$ be the actions surviving iterated elimination of strictly dominated strategies, equivalently the correlated rationalizable actions \cite{brandenburger1987rationalizability}, and let $\RatCor(\Gamma)$ be the set of profiles $\sigma$ such that, for each $i$, $\supp(\sigma_i)\subseteq\BRp_i(\nu)$ for some distribution $\nu$ on $\prod_{j\neq i}A_j^*$. Its pure profiles are exactly $\prod_iA_i^*$, and for $n=2$ it consists of the profiles in which each player uses a mixed strategy that is a best response to some belief over the opponent's rationalizable actions, that is, a rationalizable mixed strategy in the sense of Bernheim and Pearce.

\begin{theorem}[Endpoints of belief conjectures]\label{thm:endpoints}
$\RatB(\Gamma,K_n)=\NE(\Gamma)$ and $\RatB(\Gamma,\emptyset)=\RatCor(\Gamma)$. Consequently~\eqref{eq:chain} holds for every network.
\end{theorem}

\begin{proof}
The first identity is Theorem~\ref{thm:monotone}. For the second, the stages are products $\prod_iP_i^{(t)}$ by Proposition~\ref{prop:cartesian}; let $A_i^{(t)}$ be the set of actions in the supports of strategies in $P_i^{(t)}$. We show by induction on $t$ that
\begin{enumerate}[label=(\alph*)]
  \item $\delta_{a_i}\in P_i^{(t)}$ for every $a_i\in A_i^{(t)}$, and
  \item $P_i^{(t+1)}=\{\sigma_i\in P_i^{(t)}:\ \supp(\sigma_i)\subseteq\BRp_i(\nu)\ \text{for some}\ \nu\in\Delta(\textstyle\prod_{j\neq i}A_j^{(t)})\}$.
\end{enumerate}
Statement (a) holds trivially for $t=0$. Assume (a) at $t$. In the product stage, the fiber of $\sigma$ for player $i$ is $\{\sigma_i\}\times\prod_{j\neq i}P_j^{(t)}$ if $\sigma_i\in P_i^{(t)}$ and empty otherwise. The distributions on $A_{-i}$ induced by beliefs over this fiber are mixtures of product measures with factors in $P_j^{(t)}$; by (a) they include every point mass on $\prod_{j\neq i}A_j^{(t)}$, and they are supported on that set, so they form all of $\Delta(\prod_{j\neq i}A_j^{(t)})$. Together with~\eqref{eq:support}, this gives (b) at $t$. For (a) at $t+1$, let $a_i\in A_i^{(t+1)}$, say $a_i\in\supp(\sigma_i)$ with $\sigma_i\in P_i^{(t+1)}$. By (b), $a_i\in\BRp_i(\nu)$ for the belief $\nu$ that justifies $\sigma_i$, and $\delta_{a_i}\in P_i^{(t)}$ by (a) at $t$, since $a_i\in A_i^{(t+1)}\subseteq A_i^{(t)}$. Applying (b) to $\delta_{a_i}$ gives $\delta_{a_i}\in P_i^{(t+1)}$.

By (a) and (b), $A_i^{(t+1)}$ consists of the actions in $A_i^{(t)}$ that are best responses to some correlated belief over $\prod_{j\neq i}A_j^{(t)}$, which is the standard characterization of iterated elimination of strictly dominated strategies \cite{brandenburger1987rationalizability}. The sets stabilize after finitely many rounds, and by (b) the limit is $\RatCor(\Gamma)$. The chain~\eqref{eq:chain} follows from Theorem~\ref{thm:monotone}.
\end{proof}

\section{Finding a Predicted Profile}\label{sec:search}

Since $\NE(\Gamma)\subseteq\Rat(\Gamma,G)\subseteq\RatB(\Gamma,G)$, any algorithm for Nash equilibrium finds a predicted profile; the search problem is therefore in $\PPAD$ for two players and in $\FIXP$ for exact $n$-player solutions \cite{daskalakis2009complexity,etessami2010complexity}. On the empty network it is much easier: by~\eqref{eq:pure-in-mixed} and Theorems~\ref{thm:expost-exist} and~\ref{thm:expost-poly}, a predicted pure profile exists and can be computed in polynomial time. A single observation link changes this.

\begin{theorem}[One-sided monitoring]\label{thm:single-edge}
Let $n=2$ and $E=\{(1,2)\}$. For both variants,
\[
  \Rat(\Gamma,G)=\RatB(\Gamma,G)=\big\{(\sigma_1,\sigma_2):\ \sigma_2\in\pi_2(\NE(\Gamma)),\ \sigma_1\in\BR_1(\sigma_2)\big\}.
\]
Membership can be decided by one linear program, and from any predicted profile a Nash equilibrium $(\sigma'_1,\sigma_2)$ can be computed in polynomial time.
\end{theorem}

\begin{proof}
Let $\sigma\in\RatB(\Gamma,G)$. Player~1 observes everyone, so $\sigma_1\in\BR_1(\sigma_2)$. Player~2's belief $\mu$ is supported on profiles $(\hat\sigma_1,\sigma_2)\in\RatB(\Gamma,G)$; applying player~1's condition to each of them gives $\hat\sigma_1\in\BR_1(\sigma_2)$. Since $u_2$ is linear in player~1's strategy, $\sigma_2$ is a best response to the mean $\bar\sigma_1=\sum\mu(\hat\sigma)\hat\sigma_1$, and $\bar\sigma_1\in\BR_1(\sigma_2)$ because best-response sets are convex. Hence $(\bar\sigma_1,\sigma_2)$ is a Nash equilibrium. Conversely, if $(\sigma^*_1,\sigma_2)\in\NE(\Gamma)$ and $\sigma_1\in\BR_1(\sigma_2)$, then $\{(\sigma_1,\sigma_2),(\sigma^*_1,\sigma_2)\}$ is self-sustaining under point conjectures, with player~2 witnessed by $(\sigma^*_1,\sigma_2)$. Together with $\Rat\subseteq\RatB$ this proves the identity. Finally, $\sigma_2\in\pi_2(\NE(\Gamma))$ if and only if the linear system
\[
  \sigma'_1\in\Delta(A_1),\quad \sigma'_1(a_1)=0\ \ \forall a_1\notin\BRp_1(\sigma_2),\quad u_2(\sigma'_1,a_2)\ge u_2(\sigma'_1,b_2)\ \ \forall a_2\in\supp(\sigma_2),\ b_2\in A_2
\]
is feasible, and any solution completes $\sigma_2$ to an equilibrium.
\end{proof}

\begin{theorem}[Hardness on every non-empty network]\label{thm:hardness}
For every $n\ge2$ and every network $G$ with $E\neq\emptyset$, computing an element of $\Rat(\Gamma,G)$ or of $\RatB(\Gamma,G)$ is $\PPAD$-hard, even when $|A_k|\ge2$ for every player $k$. For $n=2$ the problem is $\PPAD$-complete.
\end{theorem}

\begin{proof}
Let $(i,j)\in E$ and let $\Gamma_0$ be a bimatrix game. Let $i$ and $j$ play $\Gamma_0$, and give every other player two actions that affect no one's payoff, including her own. The strategies of these players are unconstrained and irrelevant to $i$ and $j$, so the argument of Theorem~\ref{thm:single-edge} applies to the pair $(i,j)$: if $(j,i)\notin E$, the $j$-coordinate of any predicted profile is a Nash equilibrium strategy of $\Gamma_0$, and the complementary strategy of $i$ is obtained by linear programming; if $(j,i)\in E$, the restriction to $\{i,j\}$ is itself an equilibrium of $\Gamma_0$. Hardness follows from \cite{chen2009settling,daskalakis2009complexity}, and membership for $n=2$ from the reduction to Nash equilibrium.
\end{proof}

\begin{remark}[A link pins down the observed player only]\label{rem:stackelberg}
Theorem~\ref{thm:single-edge} constrains the observed player but not the observer. In the game of Table~\ref{tab:stackelberg}, the unique Nash equilibrium is $(\tfrac12T+\tfrac12B,\ \tfrac23L+\tfrac13R)$. Under $E=\{(1,2)\}$, player~2 must play $\tfrac23L+\tfrac13R$, but player~1, who is indifferent against this strategy, may play anything. For instance, $(T,\ \tfrac23L+\tfrac13R)$ is predicted although it is not an equilibrium, since player~2 would rather play $R$ against $T$; across predicted profiles, player~2's payoff ranges from $\tfrac23$ to $\tfrac73$. The link should not be read as commitment: a leader would choose her strategy in anticipation of the follower's reply, for example committing to $L$, to which player~1 strictly prefers $T$ \cite{conitzer2006computing}. Here, as in Nash equilibrium, player~2 takes player~1's strategy as given; observation gives information to the observer, not leverage to the observed.
\end{remark}

\begin{table}[htbp]
\centering
\caption{One-sided monitoring versus commitment.}
\label{tab:stackelberg}
\vspace{2pt}
\begin{tabular}{l c c}
\toprule
 & $L$ & $R$ \\
\midrule
$T$ & $2.5,\ 2$ & $0,\ 3$ \\
$B$ & $2,\ 1$ & $1,\ 0$ \\
\bottomrule
\end{tabular}
\end{table}

\section{Verification on Hierarchical Networks}\label{sec:verify}

We now ask whether a given rational profile $\sigma$ belongs to the predicted set. On $K_n$ this is a best-response check; in general it requires certifying a self-sustaining set containing $\sigma$. Suppose the players can be ordered so that each observes all players after her:
\begin{equation}\label{eq:hierarchy}
  \Nbr_i(G)=\{i+1,\dots,n\}\qquad(i=1,\dots,n).
\end{equation}
Equivalently, $G$ is a transitive tournament; equivalently, both $G$ and its complement are acyclic, since a two-cycle in either graph is excluded and an acyclic tournament is transitive. Player~$1$ observes everyone, and player~$n$ is observed by everyone and observes no one. Players with a single action can be deleted without affecting anything, and the induced network on the remaining players again satisfies~\eqref{eq:hierarchy}; we may therefore assume $|A_i|\ge2$ for all $i$.

For $\tau\in\Delta(A_n)$, let $\Gamma|_\tau$ be the $(n-1)$-player game obtained by fixing player $n$'s strategy to $\tau$, on the induced network $G'$ on $\{1,\dots,n-1\}$.

\begin{theorem}[Recursive verification]\label{thm:tournament}
Under~\eqref{eq:hierarchy}, $\sigma=(\sigma_{-n},\sigma_n)\in\RatB(\Gamma,G)$ if and only if
\begin{enumerate}[label=(\alph*)]
  \item $\sigma_{-n}\in\RatB(\Gamma|_{\sigma_n},G')$, and
  \item $\sigma_n\in\BR_n(\mu)$ for some belief $\mu$ with at most $|A_n|+1$ atoms in $\RatB(\Gamma|_{\sigma_n},G')$.
\end{enumerate}
The same holds for $\Rat$, with $\mu$ a single profile. Unfolding the recursion yields, for $\Rat$, a certificate of $2^n-n-1$ strategy vectors, linear in the size $M=n\prod_i|A_i|$ of the normal form, and, for $\RatB$, a certificate of polynomial size in $M$. Membership is therefore in $\ER$ for both variants, and in $\Poly$ for $n=2$.
\end{theorem}

\begin{proof}
Since $n\in\cNbr_i(G)$ for every $i$, all conjectures about a profile whose $n$-th coordinate is $\sigma_n$ are supported on profiles with the same $n$-th coordinate.

($\Rightarrow$) The set $\{\omega_{-n}:\omega\in\RatB(\Gamma,G),\ \omega_n=\sigma_n\}$ is self-sustaining in $(\Gamma|_{\sigma_n},G')$, because conjectures of players $i<n$ stay in it and best responses in $\Gamma|_{\sigma_n}$ coincide with those in $\Gamma$. It contains $\sigma_{-n}$ and the $(n-1)$-coordinate parts of the atoms of player $n$'s belief.

($\Leftarrow$) Let $\mathcal{S}_0\ni\sigma_{-n}$ and $\mathcal{S}_1,\dots,\mathcal{S}_k$, containing the atoms of $\mu$, be self-sustaining in $(\Gamma|_{\sigma_n},G')$, and set $\mathcal{S}=\{(\omega_{-n},\sigma_n):\omega_{-n}\in\bigcup_m\mathcal{S}_m\}$. Players $i<n$ are justified by their conjectures in $\bigcup_m\mathcal{S}_m$, extended by $\sigma_n$. Because $\Nbr_n(G)=\emptyset$, the belief $\mu$, extended by $\sigma_n$, justifies player~$n$ at every element of $\mathcal{S}$. Hence $\mathcal{S}$ is self-sustaining and contains $\sigma$.

\emph{Certificate size.} A certificate for $\sigma$ consists of the atoms of player $n$'s belief and their weights, together with certificates for $\sigma_{-n}$ and for each atom in the $(n-1)$-player game $\Gamma|_{\sigma_n}$; a one-player game requires only a best-response check. For point conjectures the number of strategy vectors satisfies $T(n)=2T(n-1)+(n-1)$, $T(1)=0$, so $T(n)=2^n-n-1$, and $M\ge n2^n$ gives $O(M)$ real entries. For belief conjectures, $T(n)\le(|A_n|+2)T(n-1)+(|A_n|+1)(n-1)$, which by induction gives $T(n)\le n\prod_i(|A_i|+2)\le n2^n\prod_i|A_i|\le M^2$. In both cases the constraints are polynomially many polynomial inequalities of size $O(M)$, which is an existential sentence over the reals of polynomial size \cite{canny1988some,schaefer2017fixed}. For $n=2$ the network is the single edge $(1,2)$, and Theorem~\ref{thm:single-edge} applies.
\end{proof}

The recursion peels off the player who is observed by all, which is possible precisely because monitoring is acyclic. The next section shows that once monitoring is cyclic, finite unfolding can fail.

\section{Cyclic Monitoring}\label{sec:cycle}

On the empty network, elimination stops after finitely many rounds (Theorem~\ref{thm:empty}), and on hierarchies every predicted profile has a finite certificate (Theorem~\ref{thm:tournament}). We show that both properties can fail on the smallest network of neither kind: the directed three-cycle $C_3$ with edges $(1,2),(2,3),(3,1)$. Indices are taken modulo $3$, so player $i$ observes $i+1$ and does not observe $i+2$.

In the games below $A_i=\{0,1\}$, a strategy is $p_i=\sigma_i(1)$, $u_i(0,\cdot)\equiv0$, and $u_i(1,\cdot)$ depends additively on $a_{i+1}$ and $a_{i+2}$. Player $i$'s gain from action $1$ is then linear in $p_{i+1}$ and in the \emph{expected} conjectured value of $p_{i+2}$. By~\eqref{eq:support}, an interior $p_i\in(0,1)$ requires this gain to be zero, which links what $i$ observes to what she must conjecture; around the cycle these links compound.

\begin{theorem}[Non-terminating elimination]\label{thm:doubling}
Let $\Gamma_{\mathrm{dbl}}$ have payoffs $u_i(1,a_{i+1},a_{i+2})=a_{i+2}-2a_{i+1}$. On $C_3$, for both variants and every $t\ge0$,
\[
  \Rat_t\cap(0,1]^3=\RatB_t\cap(0,1]^3=(0,2^{-t}]^3.
\]
Hence the stages decrease strictly at every round, elimination never terminates, and no profile in which every player uses action $1$ with positive probability is predicted.
\end{theorem}

\begin{proof}
Player $i$'s gain from action $1$ is $\Delta_i=\bar p_{i+2}-2p_{i+1}$, where $\bar p_{i+2}$ is the expected conjectured value of $p_{i+2}$. If $p_i>0$, action $1$ is in the support of $\sigma_i$, so by~\eqref{eq:support} it must be optimal: $\Delta_i\ge0$, with equality if $p_i<1$.

We prove the identity for both variants simultaneously by induction on $t$; the case $t=0$ is trivial. Assume it holds at $t$.

\emph{Upper bound.} Let $\omega\in\RatB_{t+1}\cap(0,1]^3$ and fix $i$. Since $\omega_i>0$, player $i$'s belief has mean at least $2\omega_{i+1}$ in coordinate $i+2$, so some atom $\omega'\in\RatB_t$ satisfies $\omega'_i=\omega_i>0$, $\omega'_{i+1}=\omega_{i+1}>0$ and $\omega'_{i+2}\ge2\omega_{i+1}>0$. Thus $\omega'\in\RatB_t\cap(0,1]^3=(0,2^{-t}]^3$, and $\omega_{i+1}\le\tfrac12\omega'_{i+2}\le2^{-(t+1)}$. As $i$ is arbitrary, $\omega\in(0,2^{-(t+1)}]^3$. Since $\Rat_{t+1}\subseteq\RatB_{t+1}$, the same bound holds for point conjectures.

\emph{Lower bound.} Let $\omega\in(0,2^{-(t+1)}]^3$ and fix $i$. The profile $\omega'$ with coordinates $(\omega_i,\omega_{i+1},2\omega_{i+1})$ in positions $(i,i+1,i+2)$ lies in $(0,2^{-t}]^3=\Rat_t\cap(0,1]^3$, agrees with $\omega$ on $\cNbr_i=\{i,i+1\}$, and makes player $i$ indifferent. Hence $\omega\in\Rat_{t+1}\subseteq\RatB_{t+1}$.
\end{proof}

The boundary of the strategy space survives: $(0,0,0)$ is a Nash equilibrium, and $(0,0,p_3)$ is predicted for every $p_3$, with witness set $\{(0,0,p_3),(0,0,0)\}$. What fails is finiteness of the process. Because indifference forces a doubling along each observation link, a conjecture that justifies one player imposes a stronger requirement on the next, and the requirement propagates around the cycle indefinitely. Beliefs do not help, since some atom must carry at least the mean.

Reversing the slope produces the dual phenomenon, and here the two variants part ways.

\begin{theorem}[Finite and infinite witnesses]\label{thm:halving}
Let $\Gamma_{\mathrm{half}}$ have payoffs
$u_i(1,a_{i+1},a_{i+2})=a_{i+2}-\tfrac12a_{i+1}$. Then, on $C_3$:
\begin{enumerate}[label=(\roman*)]
  \item $(0,1)^3\subseteq\Rat(\Gamma_{\mathrm{half}},C_3)\subseteq\RatB(\Gamma_{\mathrm{half}},C_3)$;
  \item under point conjectures, every self-sustaining set containing an interior profile is infinite;
  \item under belief conjectures, every $(p_1,p_2,p_3)\in(0,1)^3$ lies in the self-sustaining set $\prod_j\{0,p_j,1\}$ of $27$ profiles.
\end{enumerate}
\end{theorem}

\begin{proof}
An interior $p_i$ now requires $\bar p_{i+2}=\tfrac12p_{i+1}$.

(i) For $(p_1,p_2,p_3)\in(0,1)^3$, the interior profiles $(p_1,p_2,\tfrac12p_2)$, $(\tfrac12p_3,p_2,p_3)$ and $(p_1,\tfrac12p_1,p_3)$ match what players $1$, $2$ and $3$ observe and make them indifferent. Hence $(0,1)^3$ is self-sustaining under point conjectures.

(ii) Let $\mathcal{S}$ be self-sustaining under point conjectures and $\omega^{(0)}\in\mathcal{S}\cap(0,1)^3$. Set $k_0=1$; let $\omega^{(r+1)}\in\mathcal{S}$ be a witness for player $k_r$ at $\omega^{(r)}$ and $k_{r+1}=k_r+1$. With $v_r=\omega^{(r)}_{k_r+1}$, the witness satisfies $\omega^{(r+1)}_{k_{r+1}}=v_r$ and, by indifference of the interior player $k_r$, $\omega^{(r+1)}_{k_{r+1}+1}=\tfrac12v_r$. Inductively all these coordinates are interior and $v_r=2^{-r}v_0$. Profiles $\omega^{(r)}$ and $\omega^{(s)}$ with $r<s$ and $r\equiv s\pmod3$ differ in coordinate $k_r+1$, so $\mathcal{S}$ is infinite.

(iii) Let $\mathcal{S}=\prod_j\{0,p_j,1\}$. For $\omega\in\mathcal{S}$ and player $i$, the fiber contains profiles whose $(i+2)$-coordinate is $0$, $p_{i+2}$ or $1$. If $\omega_i$ is interior, a belief on the values $0$ and $1$ with mean $\tfrac12\omega_{i+1}\in[0,\tfrac12]$ makes $i$ indifferent. If $\omega_i=0$, the point conjecture with value $0$ gives $\Delta_i\le0$; if $\omega_i=1$, the point conjecture with value $1$ gives $\Delta_i\ge0$.
\end{proof}

Together with Theorem~\ref{thm:tournament}, these games separate acyclic from cyclic monitoring and illustrate the role of the conjecture format. On hierarchies, finite certificates always exist. On $C_3$, elimination may never stop under either variant. In $\Gamma_{\mathrm{half}}$, point conjectures require infinite witness sets, because a single hypothesis must be refined along an infinite chain, whereas beliefs short-circuit the chain by mixing boundary profiles. Whether belief conjectures always admit finite witnesses is open.

\section{Ex-Post Monitoring}\label{sec:expost}

When players observe the realized actions of their neighbors rather than their strategies, the analogue of Definition~\ref{def:rat} lives on pure profiles:
\[
  \RatP(\Gamma,G)=\bigcup\Big\{\mathcal{S}\subseteq A:\ \forall a\in\mathcal{S}\ \forall i\ \exists a'\in\mathcal{S}\ \text{with}\ a'_{\cNbr_i(G)}=a_{\cNbr_i(G)},\ a_i\in\BRp_i(a'_{-i})\Big\}.
\]
On $K_n$ this is the set of pure Nash equilibria. On the empty network it is the set of profiles surviving iterated elimination of actions that are never best responses to a pure profile of surviving opponent actions. This can exclude actions used in mixed equilibria: in the game of Table~\ref{tab:point-pure}, $M$ is never a best response to a pure action of the column player, yet $(M,\tfrac12L+\tfrac12R)$ is a Nash equilibrium.

\begin{table}[htbp]
\centering
\caption{$M$ is a best response only to non-degenerate mixtures.}
\label{tab:point-pure}
\vspace{2pt}
\begin{tabular}{l c c}
\toprule
 & $L$ & $R$ \\
\midrule
$T$ & $0,\ 1$ & $1,\ 0$ \\
$M$ & $0.6,\ 0$ & $0.6,\ 0$ \\
$B$ & $1,\ 0$ & $0,\ 1$ \\
\bottomrule
\end{tabular}
\end{table}

\begin{theorem}[Existence dichotomy]\label{thm:expost-exist}
If $E=\emptyset$, then $\RatP(\Gamma,G)\neq\emptyset$ for every finite game. If $E\neq\emptyset$, then for every $n\ge2$ there is an $n$-player game with $\RatP(\Gamma,G)=\emptyset$.
\end{theorem}

\begin{proof}
If $E=\emptyset$, the stages are products $\prod_iA_i^{(t)}$, where $A_i^{(t+1)}$ contains the actions of $A_i^{(t)}$ that are best responses to some profile in $\prod_{j\neq i}A_j^{(t)}$. Each $A_i^{(t+1)}$ is non-empty: a best response to any $a_{-i}\in\prod_{j\neq i}A_j^{(t)}$ is, a fortiori, a best response to a profile in $\prod_{j\neq i}A_j^{(s)}$ for every $s\le t$, and hence lies in $A_i^{(t+1)}$. The decreasing sequence is finite, so its limit is a non-empty product.

If $(i,j)\in E$, let $i$ and $j$ play Matching Pennies and give all other players a single action. If $(j,i)\in E$, predicted profiles would be pure equilibria, which do not exist. Otherwise, as in Theorem~\ref{thm:single-edge}, player $j$'s witness $a'$ satisfies $a'_j=a_j$, $a_j\in\BRp_j(a'_i)$ and $a'_i\in\BRp_i(a_j)$, so $(a'_i,a_j)$ would be a pure equilibrium.
\end{proof}

\begin{theorem}[Polynomial-time computation]\label{thm:expost-poly}
For every game and network, $\RatP(\Gamma,G)$ can be computed in time $O(n|A|^2)$.
\end{theorem}

\begin{proof}
By the self-witnessing property, $a\in\mathcal{S}$ has a witness for $i$ in $\mathcal{S}$ if and only if some $a'\in\mathcal{S}$ with $a'_{\cNbr_i}=a_{\cNbr_i}$ satisfies $a'_i\in\BRp_i(a'_{-i})$, that is, has player $i$ best-responding within $a'$ itself. Precompute in time $O(n|A|)$ the indicators $b_i(a)=\mathbb{I}[a_i\in\BRp_i(a_{-i})]$. Starting from $\mathcal{S}=A$, each round computes, for every $i$, the set $K_i$ of keys $a'_{\cNbr_i}$ of profiles $a'\in\mathcal{S}$ with $b_i(a')=1$, and keeps the profiles $a$ with $a_{\cNbr_i}\in K_i$ for all $i$. Indexing keys by an array of size at most $|A|$, a round takes $O(n|A|)$ time. This is greatest-fixed-point iteration on the finite lattice $2^A$; every round except the last removes a profile, so there are at most $|A|$ rounds.
\end{proof}

\section{Conclusion}

Network rationalizability treats the observation structure of a game as a primitive and asks what common knowledge of rationality implies when each player sees only part of what others do. A single observation link transmits the full computational hardness of Nash equilibrium; hierarchical monitoring admits compact certificates; and cyclic monitoring can produce elimination processes that never stop. Comparing point and belief conjectures separates the effects of the network from those of the conjecture format: non-termination on the cycle persists under beliefs, whereas in our example infinite witnesses do not.

Several questions remain open. First, is membership decidable on every network? Remark~\ref{rem:cells} reduces the operators to finitely many polynomial rules, but Theorem~\ref{thm:doubling} rules out termination of naive elimination; relatedly, it is open whether the predicted sets are always semi-algebraic. Second, whether belief conjectures always admit finite witness sets, in contrast to Theorem~\ref{thm:halving}(ii), is open. Third, for two players on the empty network our algorithm for point-rationalizability is exponential only in the number of actions; whether membership is polynomial is open. Fourth, a type-space characterization in the spirit of \cite{tan1988bayesian,aumann1995epistemic} would make precise the epistemic reading of self-sustaining sets. Finally, adding links shrinks the set of predictions, but, as Remark~\ref{rem:stackelberg} shows, it does so asymmetrically, constraining the observed player while leaving the observer free among her best responses. How such asymmetric constraints translate into welfare, and in particular when being observed helps or hurts a player, is a natural question for information and network design.

\end{document}